\documentclass[aps,prb,twocolumn,letterpaper]{revtex4}
\usepackage{graphicx}
\usepackage{dcolumn}
\usepackage{amsmath}
\usepackage{amssymb}
\usepackage[caption=false]{subfig}
\usepackage{soul}
\usepackage{hyperref}
\usepackage[usenames,dvipsnames]{color}
\definecolor{orange}{rgb}{1,0.5,0}
\usepackage{float}
\usepackage{epstopdf}
\usepackage{ulem}

\def\kv{{\bf k}}

\def\beq{\begin{equation}}
\def\eeq{\end{equation}}

\def\beqa{\begin{eqnarray}}
\def\eeqa{\end{eqnarray}}

\begin{document}

\title{Analytic Expression for the Entanglement Entropy of a 2D Topological Superconductor}
\author{Jan Borchmann and T. Pereg-Barnea}
\affiliation{Department of Physics and the Centre for Physics of Materials, McGill University, Montreal, Quebec,
Canada H3A 2T8}
\date{\today}
\begin{abstract}
We study a model of two dimensional, topological superconductivity on a square lattice.  The model contains hopping, spin orbit coupling and a time reversal symmetry breaking Zeeman term.  This term, together with the chemical potential act as knobs that induce transitions between trivial and topological superconductivity.  
As previously found numerically, the transitions are seen in the entanglement entropy as cusps as a function of model parameters.  In this work we study the entanglement entropy analytically by keeping only its most important components.  Our study is based on the intuition that the number of Fermi surfaces in the system controls the topological invariant.  With our approximate expression for the entanglement entropy we are able to extract the divergent entanglement entropy derivative close to the phase transition.
\end{abstract}
\maketitle
\section{Introduction}
Entanglement\cite{EPR,schroedinger1,schroedinger2} is one of the most fascinating fundamental aspects of quantum systems that has no classical equivalent. 
Its most straight forward demonstration is through thought experiments on a system of few particles.  Nevertheless, in recent years the degree of entanglement in large systems, measured through its entanglement entropy (EE), has become a standard tool in characterizing many body systems.\cite{amico,horodecki}.  More specifically, the EE (or bipartite entanglement) is found by dividing the system into two parts, $A$ and $B$.  The reduced density matrix of subsystem $A$ is defined as the result of tracing out the degrees of freedom associated with subsystem $B$ in the density matrix.  The logarithm of the reduced density matrix is then used to define the entanglement entropy.  This, in general, leads to two categories of systems: entangled or separable.  If the two subsystems are not entangled the density matrix of the full system contain two separate blocks corresponding to each of the subsystems. In this case the system is called separable and contains no entanglement. If this is not the case, the system is entangled.  

The entanglement entropy has been shown to be a sensitive indicator of the topology in systems with intrinsic topological order\cite{Wildeboer} and recently it has also been studied in the context of symmetry protected topological states.  It has been studied numerically in the clean limit\cite{oliveira, ding, bray-ali, borchmann1} as well as in disordered systems\cite{mondragon,andrade,pouranvari,mondragon2,vijay,borchmann2}, critical systems\cite{hsiehEE} and topological states\cite{prodan, gilbert, brown}. Furthermore phase transitions in gapless states have been examined via the EE\cite{rodney}. It should be noted, however, that in general the EE is not measurable except for certain, well designed situations\cite{klich,cardy,abanin,elliott,thomas}. It is usually used as a theoretical tool for characterizing model systems.  In particular, the EE of symmetry protected topological states has been shown to exhibit a cusp as a function of model parameters when the parameters are tuned across a topological phase transition\cite{borchmann1,borchmann2}.  This is the focus of the current manuscript. 

Topological insulators and superconductors are normally characterized by topological invariants (a Chern number or Z$_2$-invariant, depending on the class). In non-interacting, clean systems these invariants can be easily computed using the Berry curvature while in interacting systems the calculation requires knowing the full Green's function and therefore the full spectrum. 
However, the full spectrum of large interacting systems is usually not accessible as most numerical methods are geared toward finding the ground state or a thermally averaged energy.
In this respect the EE may prove useful as it can be defined using the ground state alone.  This motivates our study of EE in symmetry protected topological systems.

There are several contributions to the entanglement entropy. For two dimensional free electron systems it has been shown that the entanglement entropy has the form 
\begin{equation}
S = \alpha L - \gamma + \dots,
\end{equation}
where $L$ is the cross section of the partition used to calculate the entanglement entropy and $\dots$ stands for subleading terms. The fact that the entanglement entropy is proportional to the cross section of the partition instead of the volume is called the area law\cite{eisert} where 'area' refers to the size of the boundary of the partition, in two dimensions this is a length. The term $\gamma$ is called topological entanglement entropy.  In systems with intrinsic topological order the entanglement entropy acquires this contribution, which equals the logarithm of the total quantum dimension of the system\cite{kitaev,Levin2}.  This term is not applicable for symmetry protected states (SPTs) and is strictly zero in topological insulators and superconductors.

Additional terms can arise due to corners in the partition\cite{oliveira}. This can be understood as follows.  In a finite partition, call it $A$, of size comparable to or smaller than the correlation length, the entanglement entropy is enhanced due to correlations between states in the other partition, $B$, on different sides of the partition. In a large partition with corners, close to the corner of partition $A$ there could be two states of subsystem $B$ which are closer than the correlation length.  The path connecting them goes through subsystem $A$ and affects the entanglement entropy. This leads to additional terms in the entanglement entropy provide subleading corrections to the area law.

\begin{figure}[t]
\begin{center}
\includegraphics[width = 0.75\linewidth]{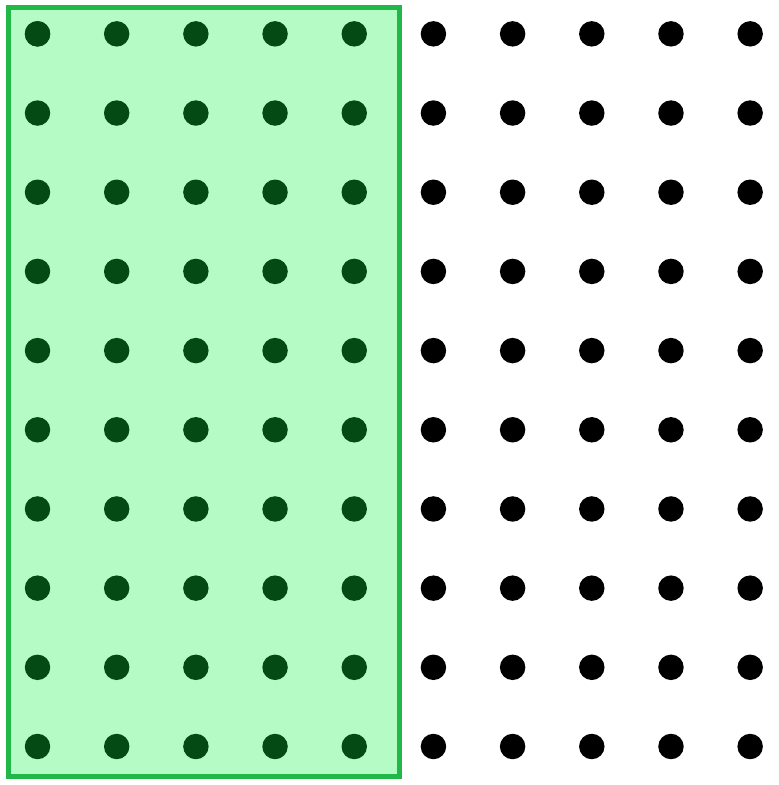}
\end{center}
\caption{The cornerless partition with the left (right) part being subsystem $A$ ($B$)\label{partition}}
\end{figure}

The reduced density matrix as well as the entanglement entropy are often difficult to evaluate analytically and are only accessible numerically. This is due to the fact that the calculation includes large matrices with non-generic features, making it hard to find general solutions. Additionally, partitioning the system breaks translation invariance in one or more directions, requiring a real space treatment. Moreover, the specific shape of the partition may also have contributions to the entanglement entropy.
We therefore adopt a corner-less partition and avoid the effects of corners.  The system is divided into left and right subsystems, $A$ and $B$, respectively, and consequently preserves translation invariance in one spatial direction, this is depicted in Fig.~\ref{partition}.

The model we study is inspired by proposals for realizing two-dimensional topological superconductors in heterostructures\cite{Fu,Sau,Alicea}. In these heterostructures different layers provide the following essential ingredients needed to realize the topological superconductor.  One or more layers provide non-trivial topology through spin-orbit coupling and a Zeeman field and an additional layer provides pairing through the proximity effect. The combination of these ingredients leads to chiral $p$-wave pairing in the valance band.

The paper is structured as follows: In section \ref{from_s_to_p}, starting out with a spin-orbit coupled $s$-wave superconductor, we derive the effective $p$-wave model in the large Zeeman limit. In section \ref{ee} we derive the expression for the entanglement entropy and analyze the correlation functions of the system.  We conclude in section \ref{conclusion}.

\section{Model Hamiltonian}\label{from_s_to_p}
We describe our system via the Hamiltonian\cite{Alicea,Sau} 
\begin{equation}
H=T+H_{SC}+H_{SO}.
\end{equation}
With the kinetic term,
\begin{equation}
T = -t\sum_{\langle i,j\rangle,\sigma} c_{i,\sigma}^\dagger c_{j,\sigma} - \mu \sum_i c_{i,\sigma}^\dagger c_{i,\sigma}
\end{equation}
which includes nearest neighbour hopping on a square lattice as well as the chemical potential, $\mu$. The second term introduces the pairing and reads,
\beq
H_{SC} = \sum_{\kv}\left( \Delta_{s} c_{\kv,\uparrow} c_{-\kv,\downarrow} +\text{h.c.}\right),
\eeq
where $\Delta_{s}$ is a superconducting $s$-wave order parameter. The last term includes the spin-orbit coupling and Zeeman field,
\begin{equation}
H_{SO} = \sum_{\kv} \Psi_{\kv}^\dagger \mathcal{H}_\kv \Psi_{\kv},
\end{equation}
with  $\Psi_{\kv}=(c_{\kv,\uparrow}, c_{\kv,\downarrow})^T$, $\mathcal{H}_\kv = {\bf d}_\kv \cdot \vec{ \sigma}$, where the Pauli matrices $\vec\sigma$ act on the spin degree of freedom and ${\bf d}_\kv=(A\sin{k_x},A\sin{k_y},M)$.  Here $A$ and $M$ represent the Rashba spin-orbit coupling and Zeeman strength, respectively.

Starting from the tight binding model with Rashba spin-orbit coupling and no pairing, one finds that this coupling has the effect of aligning the spin of the electrons in the plane orthogonal to their momentum, leading to a Dirac cone at the gamma point. Introducing a finite Zeeman coupling gaps out the Dirac point.  When the Zeeman mass $M$ is larger than the band width (determined by $t$ and $A$) the two spin-orbit coupled bands do not overlap.  The chemical potential then determines which band contributes to superconductivity.  In this work we first focus on the regime $M>4t+\mu$ where the Fermi level crosses only the lower band. We then discuss other possible cases.

In spin orbit coupled bands it is often convenient to work in a band basis, rather than a spin basis.  We therefore introduce creation/annihilation operatore for electrons in the upper and lower bands, $\Psi_\pm$ and $\Psi_\pm^\dagger$ and write 

\begin{align}
\begin{split}
\Psi_{\kv} = \phi_- (\kv) \Psi_- (\kv) + \phi_+ (\kv) \Psi_+ (\kv).
\end{split}
\end{align}
where $\phi_\pm(\kv)$ are scalar function representing the basis transformation.
In the absence of pairing this transformation diagonalizes the kinetic part of the Hamiltonian, $H_0 = T + H_{SO}$ and leads to the following dispersion: 
\begin{align}
\begin{split}
\epsilon_{\pm} = &-2t(\cos{k_x} + \cos{k_y}) - \mu \\ &\pm \sqrt{A^2(\sin^2{k_x} + \sin^2{k_y}) + M^2}.
\end{split}
\end{align}
In this basis the pairing part of the Hamiltonian reads
\begin{align}
\begin{split}
H_{SC} = \sum_{\kv}  &\left[ \Delta_{+-}(\kv) \psi^\dagger_+ (\kv) \psi^\dagger_- (-\kv) \right.  \\&\left. +\Delta_{--} (\kv) \psi_-^\dagger (\kv) \psi_-^\dagger (-\kv) \right.  \\ &\left.  + \ \Delta_{++} (\kv) \psi^\dagger_+ (\kv) \psi_+ (-\kv) + \text{h.c.} \right].
\end{split}
\end{align}
Here, $ \Delta_{+-}$  denotes an interband pairing function of $s$-wave symmetry; the other two terms are intraband pairings of $p$-wave symmetry. The intraband pairing is given by
\begin{align}
\begin{split}
 \Delta_{--} (\kv) = \frac{A\Delta_s(\sin{k_y}-i\sin{k_x})}{2\sqrt{M^2 + A^2 (\sin^2{k_x} + \sin^2{k_y})}}= \Delta_{++}^*(\kv)
\end{split}
\end{align}

As shown previously\cite{Alicea,farrell1,borchmann1}, this model exhibits a topological phase transition, when varying the parameters of the Hamiltonian.  This can be seen by calculating the Chern number.  More intuitively, one sees that a topological phase arises when there is only one spin-orbit coupled band which participates in the pairing. For fixed $\mu,A$ and $\Delta_s$, one can show that the phase transition takes place at $M_\pm = \sqrt{\Delta_s^2 + (\pm 4 - \mu)^2}$
where the topological phase is for $M \in (M_- , M_+)$. 

It can be shown that in the large Zeeman coupling regime and in the limit of small order parameter $\Delta_s$, the interband $\Delta_{+-}$ pairing can be neglected. Thus, we can project out the upper band altogether and arrive at a chiral $p$-wave model:
\begin{align}
\begin{split}
H = \sum_{\kv} \left[ \epsilon_- (\kv) \psi^\dagger_- \psi_- + \Delta_{--} (\kv) \psi^\dagger_- (\kv) \psi^\dagger_- (-\kv) + \text{h.c.}  \right],
\end{split}
\end{align}
where in this limit we may approximate
\begin{align}
\begin{split}
\Delta_{--} (\kv) \approx \frac{A\Delta}{2|M|}(-i\sin{k_x} + \sin{k_y}).
\end{split}
\end{align}
We therefore drop the subscripts and arrive at an effective spinless model. 

As mentioned above the system partitioning breaks translation invariance in the $x$-direction.  We therefore introduce a mixed real- and momentum-space representation, $c_{k_x k_y} = \frac{1}{\sqrt{N}}\sum_{i_x}e^{-ir_i^x k_x}c_{i_x k_y}$.  Our model can therefore be regarded as a collection of chains in the $x$ direction with $k_y$ controlling the chain parameters. The kinetic part of the Hamiltonian, $H_0 = T+H_{SO}$, is given by:
\begin{align}
\begin{split}
H_0 = \sum_{k_y}&\left[ \sum_{i_x}(-2t\cos{k_y} - \mu - |M|)c_{i_x k_y}^\dagger c_{i_x k_y}\right. \\  &\left.  - t (c_{i_x k_y}^\dagger c_{i_x+1 k_y} + \text{h.c.})\right] ,
\end{split}
\end{align}
where we have ignored the parameter $A$. The pairing is,
\begin{align}
\begin{split}
H_{SC} = \alpha\sum_{k_y,i_x}&\left[ - c^\dagger_{i_x+1, k_y}c^\dagger_{i_x,-k_y} \right. \\ &\left. + \sin{k_y}\ c^\dagger_{i_x,k_y}c^\dagger_{i_x,-k_y} + \text{h.c.}\right],
\end{split}
\end{align}
where $\alpha =  \frac{A\Delta}{2M}$. 
We therefore arrived at an effective spinless Hamiltonian with $M \gg \Delta,A,t$. As a consistency check, we examine the topological properties of the effective Hamiltonian compared to the full model. For the full spin-orbit coupled $s$-wave superconductor the large $M$ phase transition from a topological phase to the trivial phase takes place at $M_+ = \sqrt{\Delta_s^2 + (4-\mu)^2}$. For the effective model the phase transition takes place at $|-\mu-M| = 4$. Thus, for small $\Delta_s$ our approximation will reproduce the behaviour of the full system reasonably well.

Due to translation symmetry in the $y$-direction it is possible to treat the Hamiltonian $H$ as a sum of one-dimensional Hamiltonians with a parameter $k_y$. Thus, when partitioning the system, we can think of it as cutting each one-dimensional chain into two parts. 

Another way to look at our current model is as follows.  Each of the $k_y$-dependent chains in our system is a one dimensional Kitaev model\cite{kitaev_model}.  Depending on its parameters the chain could be in the strong coupling trivial phase of the weak coupling topological phase.  We find that our full system is topological as long as {\it some} of the chains are in the topological regime.  Therefore, a phase transition from a trivial to a topological state occurs as soon as one chain becomes topological.  Conversely, a phase transition from a topological to a trivial state occurs when all chains become trivial.  

\section{Calculation of the Entanglement Entropy}\label{ee}
The first step in calculating the EE is to define the reduced density matrix $\rho_a$ by integrating out the degrees of freedom associated with subsystem $B$. The reduced density matrix can then be used to define the entanglement Hamiltonian $H_A$ via $\rho_A = \frac{e^{-H_A}}{Z_A}$, where $Z_A$ is the partition function with respect to $\rho_A$. The eigenvalues of $H_A$ are the entanglement spectrum.  We denote these 'energies' by $E_i$, and use them to calculate the EE.  Moreover, it has been shown\cite{peschel1,peschel2} that the entanglement spectrum is related to the eigenvalues of the correlation matrix $G$, which is defined as:  
\begin{align}
\begin{split}
G = \left(  \begin{matrix}
      \langle c_{i} c^\dagger_{j} \rangle &\langle c_{i}c_{j} \rangle  \\
      \langle c_{i}^\dagger c^\dagger_{j} \rangle & \langle c_{i}^\dagger c_{j} \rangle \\
   \end{matrix}\right),
\end{split}
\end{align}
where each term is a matrix and the indices $i$ and $j$ run over the sites of subsystem $A$, the left part of our system of length $L_x$. The averages are calculated with respect to the ground state $\psi$. Denoting the eigenvalues of $G$ by $\zeta_i$, the entanglement spectrum levels are given by $E_i = \ln(\frac{\zeta_i}{1-\zeta_i})$, and therefore the EE can be written as
\begin{equation}
S = \sum_i \left[\zeta_i\log\zeta_i+(1-\zeta_i)\log(1-\zeta_i)\right]
\end{equation}
The general structure of the eigenvalues of $G$ consists of eigenvalues very close to 0 and 1 reflecting the fact that most bulk states are almost completely localized either in subsystem $A$ or $B$ and consequently do not contribute to the entanglement entropy. Intermediate values of $\zeta_i$ are caused by states that are entangled across the partition boundary and thus contribute the most to the EE.

Diagonalizing the matrix $G$ was done numerically in Refs.~[\onlinecite{borchmann1,borchmann2}] and the Chern number of the same model was calculated explicitly in Ref.~[\onlinecite{farrell1}].  In these works it was seen that the Chern number changes whenever a Fermi surface appears or shrinks to a point and disappear.  We therefore test this assumption by stripping the correlation matrix $G$ of any ingredients which are unnecessary for detecting the phase transition.  As we shall see shortly the crude approximations we make lead to an analytic expression for the EE which mimics the numerical one around the topological phase transition.

We begin by ignoring any off-diagonal (anomalous) terms in the correlation matrix due to $\alpha \ll 1$, and are therefore left with the usual particle-hole correlation.  Of course, with the off-diagonal piece ignored, there is no need to keep the Nambu notation and one can focus on terms like $\langle c_\alpha c_\beta^\dagger\rangle$.  We are therefore left with on-site and longer range correlations. For the regime we are investigating these long range correlations decay rapidly and thus it is reasonable to ignore higher order correlations.  Hence, diagonalizing such a matrix is analogous to finding the eigenstates of a tight binding model in one dimension with open boundary conditions\cite{jullien}. However, we find that including only the on-site correlations and ignoring even nearest neighbour ones is easier and sufficient in our case.  Of course, this is a good approximation only when the correlation length is not too long. A second assumption we make is that the system is large enough and the correlations die off quickly such that the correlation functions are position independent.

We are left with evaluating the onsite correlation for each $k_y$ dependent chain, $\langle c_{i_x}(k_y) c_{i_x}(k_y)^\dagger\rangle$.  Since we've ignored superconductivity this amounts to summing all of the occupation numbers $\langle c_{k} c_{k}^\dagger\rangle = 1-n_k$ over the $k_x$ momentum.  At zero temperature, this amounts to the fraction of a $2\pi$ long line, along the $k_x$ direction in the Brillouin zone, which contains (un)occupied states.  In other words, if we draw the Fermi surface in the Brillouin zone and draw a line at a specific $k_y$, what fraction of this line is (outside)inside the Fermi surface. The answer is given by
\begin{eqnarray}
\zeta_{k_y}=\langle c_{i_x}(k_y) c_{i_x}(k_y)^\dagger\rangle = \int \langle c_{k} c_{k}^\dagger\rangle dk_x = \nonumber \\
\begin{cases} 
{1 - 2k_x^0(k_y)\over 2\pi} & {\rm  particle-like~Fermi~surface}\\
{2k_x^0(k_y)\over 2\pi} & {\rm hole-like~Fermi~surface}
\end{cases} 
\end{eqnarray}
where $k_x^0(k_y)$ is the $x$-component of the Fermi vector when the $y$-component is given by $k_y$.  With our quadratic lattice dispersion we get:
\begin{equation}
k_x^0(k_y)=\arccos\left(\cos(k_y) - \frac{\mu+ |M| }{2t}\right) 
\end{equation}
which is only defined for $k_y$ values where there is a real solution (otherwise the contribution to the EE vanishes).

Putting all of the above together we are now able to write an expression for the entanglement entropy as a sum of the EEs of each chain:
\begin{eqnarray}
S = \sum_{k_y} S(k_y) \approx \frac{L_y}{2\pi}\int_0^{2\pi}\! \! \! dk_y \ S(k_y), \\
S(k_y) = \zeta_{k_y} \ln\zeta_{k_y} + (1 - \zeta_{k_y}) \ln(1-\zeta_{k_y}).
\end{eqnarray}
The first line above clearly shows the area law as the EE explicitly depends on the length of the partition, $L_y$, which is the number of sites along the $y$ direction.  Together with the second line this expression is not yet a closed form but can be evaluated easily in simple cases.

Another simplification comes from the fact that transitions happen when Fermi surfaces appear and disappear.  This amounts to the Fermi surface passing through the center or the corner of the Brillouin zone.  We can therefore replace the sum over the $k_y$ momentum by these points only and define $a = \frac{1}{\pi}\Re[\arccos(1-{\mu + |M| \over 2t})]$ and $b =  \frac{1}{\pi}\Re[\arccos(-1-{\mu + |M| \over 2t})]$.  This reduces the EE to:
\begin{align}
\begin{split}
S =& -L\left[(1-a)\left(a\ln a + (1-a)\ln(1-a)\right) \right. \\
&\left. + (1-b)\left(b\ln b + (1-b)\ln(1-b)\right)\right].\label{eq:AnalyticEE}
\end{split}
\end{align}
In this expression we have two contributions, the $a$-term from a possible phase transition at $k_y=\pi$ and the $b$-term from the $\Gamma$-point.

\begin{figure*}[t]
\begin{center}
\subfloat[~]{\includegraphics[width = 0.5\linewidth]{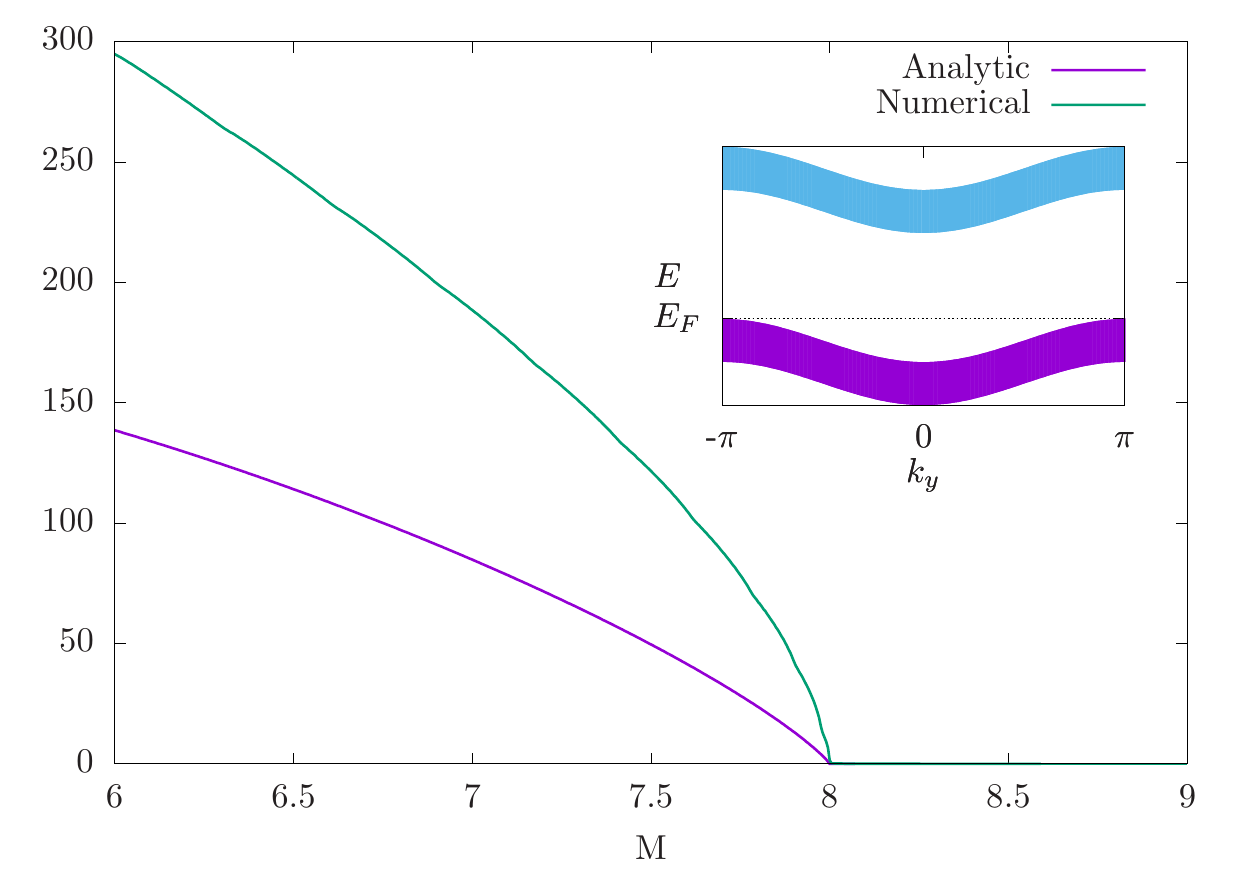}}
\subfloat[~]{\includegraphics[width = 0.5\linewidth]{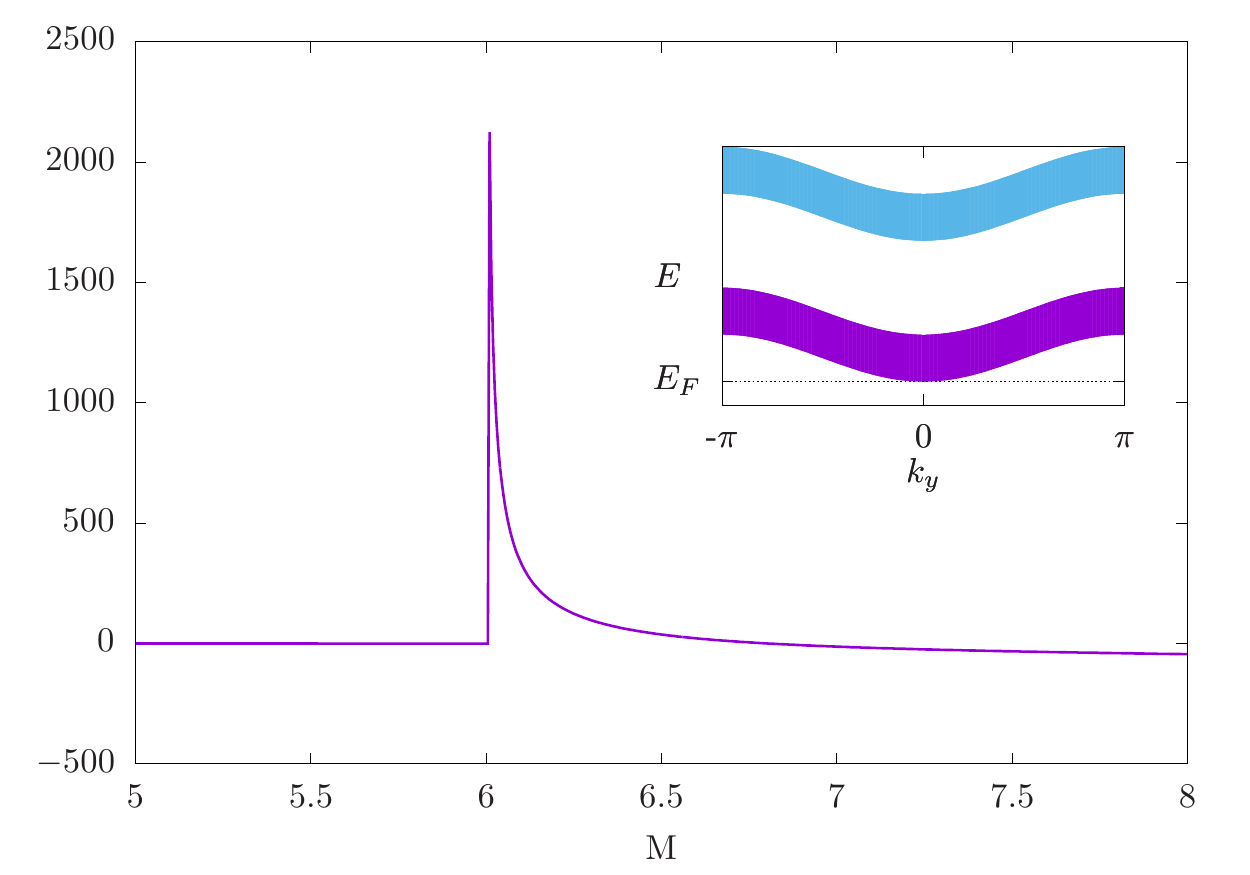}}
\end{center}
\caption{(a)Entanglement entropy $S$ for $\mu=-4t, A=0.25t$ and $L_y = 400$ calculated numerically (green) as well as analytically (red). In the inset we show the position of the two bands with respect to the Fermi energy of the system without superconductivity. (b) Derivative of the entanglement entropy with respect to $M$ for $\mu=-10t$. Again, the inset shows the band spectrum with respect to the Fermi energy. Note that the two panels represent different phase transitions with similar behaviour of the EE and its derivative\label{S}}
\end{figure*}

We can now compare the entanglement entropy calculated via Eq.~(\ref{eq:AnalyticEE}) with the exact numerical value calculated by diagonalizing the full correlation matrix $G$. As can be seen in FIG. \ref{S}(a), the analytic formula underestimates the the entanglement entropy by a factor of $O(1)$, which is not surprising due to the way we approximated the correlation matrix. Nonetheless we can see that the formula captures very well the qualitative behaviour of the EE and in particular its behaviour near the phase transition. 

As we have shown previously\cite{borchmann1,borchmann2} the EE is indeed sensitive to the topological phase transitions.  With the simple, approximate expression above we see that the topological phase transition is indeed governed by the appearance and disappearance of Fermi surfaces.  At the transition the EE has a cusp and a singularity in its derivative:
\begin{align}
\begin{split}
\frac{\partial S}{\partial M} = &\Re\left[\frac{L_y\left[ 2(1-a)\ln{(1-a)} + (2a-1)\ln{(a)}\right]}{\pi \sqrt{(|M|+\mu)(4t-|M|-\mu)}}\right. \\
&\left.  +\ (a\rightarrow b)\right. \Bigg]
\end{split}
\end{align}
From this one can immediately see the change in behaviour at the topological phase transition, where the derivative jumps to zero as shown in FIG.~\ref{S}(b).  

With the intuition about the topological phase transitions from the approximate, single band model we can now relax our single band requirement slightly.  Including the model's upper band has two consequences: (i) more Fermi surfaces and therefore more transitions may occur (ii) the singlet, inter-band pairing may not be negligible.  Therefore, if we allow $M$ to be smaller we might view more phase transitions. The inter-band pairing is relevant when there is an over lap in energy between the two bands and in particular when the Fermi surface lies in this overlap.  If we choose to ignore this inter-band pairing $\Delta_{+-}$ we may assign a 'topological' label to a trivial superconductor.  However, as there is an even number of Fermi surfaces in this case, the system will have an even Chern number and an even number of Majorana branches on each edge.  In this respect it is equivalent to a trivial superconductor.  We therefore extend our analysis by relaxing the constraint over $M$ while still ignoring interband pairing.  The EE in this case is simply the sum of EE of the two bands $S=S_- + S_+$, while the upper band EE is given by Eq.~(\ref{eq:AnalyticEE}) with $|M|\to-|M|$. Accordingly, the derivative receives a second term,
\begin{align}
\begin{split}
\frac{\partial S_+}{\partial M} = &\Re \left[ \frac{L_y\left[ 2(1-b)\ln{(1-b)} + (2b-1)\ln{(b)}\right]}{\pi \sqrt{(-|M|+\mu)(-4t+|M|-\mu)}} \right. \\
&\left.  +\ (b\rightarrow a)\right. \Bigg]
\end{split}
\end{align} 

Comparing the analytic and numeric evaluation of the entanglement entropy of our model we see that all phase transitions are captured as a singularity in $\partial S/\partial M$.  The cusps in the EE occur in places where the Chern number of the system changes, i.e., a topological phase transition.  However, there are some additional points where the EE exhibits a cusp but there is no phase transition.  This happens when the pairity of two time reversal invariant momentum (TRIM) points change simultaneously as a result of a lattice symmetry.  If this symmetry is lifted, these points in parameter space will become phase transition points.

\section{Conclusion}\label{conclusion}
In this work we have analytically calculated the entanglement entropy for a spin-orbit coupled superconductor in the large Zeeman coupling limit. In this regime the spectrum has a large gap, even without any pairing. Looking at the low energy part of the entanglement spectrum, one arrives at an effective $p$-wave superconductor. We are able to show explicitly that the entanglement entropy obeys the area law as expected. When comparing with the exact EE calculated through numerical diagonalization of the correlation matrix one finds that both indicate the same phase transitions. The derived formula is in qualitative agreement with the numerical evaluation.

The above calculation is enabled by crude approximations which relay on the pairing being small compared to the bandwidth and predominantly in the intra-band channel.  The intuition behind these approximations comes from the understanding that the topology of the superconductor is inherited from the spin winding in each spin-orbit coupled bands and depends crucially on the number of Fermi surfaces.
\section{Acknowledgments}
The authors are grateful for useful discussions with A.~Farrell and O.~Motrunich.  Financial support for this work has been provided by the Alexander McFee award (JB), NSERC and FQRNT (JB and TPB).

\bibliographystyle{aipnum4-1}
\bibliography{topoSC}

\end{document}